\documentclass[
reprint,
superscriptaddress,
showpacs,
preprintnumbers,
nofootinbib,
nobibnotes,
amsmath,
amssymb, 
aps,
prd,
floatfix
]{revtex4-1}
\pdfoutput=1 
\usepackage[utf8]{inputenc} 
\usepackage{soul}
\usepackage{graphicx}  % needed for figures
\usepackage{dcolumn}   % needed for some tables
\usepackage{bm,relsize}        % for math
\usepackage{amssymb,amsmath,mathrsfs}
\usepackage{textcomp}
\usepackage{wasysym}
\usepackage{slashed}
\usepackage{multirow}
\usepackage{lipsum, color}
\usepackage[colorlinks=true,allcolors=purple]{hyperref}
\usepackage[usenames,dvipsnames,svgnames]{xcolor}
\usepackage{booktabs}
\usepackage{xfrac}
\usepackage[capitalize]{cleveref}
\usepackage{isotope}
\usepackage{accents}
\usepackage{comment}
\usepackage{physics}
\usepackage{titlesec}

\def\ltap{\ \raise.3ex\hbox{$<$\kern-.75em\lower1ex\hbox{$\sim$}}\ }
\def\gtap{\ \raise.3ex\hbox{$>$\kern-.75em\lower1ex\hbox{$\sim$}}\ }
\def\lsim{\ \raise.3ex\hbox{$<$\kern-.75em\lower1ex\hbox{$\sim$}}\ }
\def\gsim{\ \raise.3ex\hbox{$>$\kern-.75em\lower1ex\hbox{$\sim$}}\ }
\def\beq{\begin{equation}}
\def\eeq{\end{equation}}
 \def\be{\begin{equation}} \def\ee{\end{equation}}
\def\bea{\begin{eqnarray}} \def\eea{\end{eqnarray}}

\renewcommand{\Tr}{{\rm Tr}}

\newcommand{\epsilonmix}{\epsilon_{nn'}}
\newcommand{\nprime}{{n'}}
\newcommand{\pprime}{{p'}}

\usepackage{fontawesome5} % for code link icons % 
\definecolor{blue-violet}{rgb}{0.33, 0.17, 0.89}

\titlespacing{\subsection}{0pt}{0.1in}{0.1in}

\begin{document}

%%%%%%%%%%%%%%%%%%%%%%%%%%%%%%%%%%%%%%%%%%%%
\title{Neutron portal to ultra-high-energy neutrinos}
%%%%%%%%%%%%%%%%%%%%%%%%%%%%%%%%%%%%%%%%%%%%

\author{Gustavo F. S. Alves}
\email{gustavo.figueiredo.alves@usp.br}
\affiliation{Instituto de F\'isica, Universidade de S\~ao Paulo, C.P. 66.318, 05315-970 S\~ao Paulo, Brazil}

\author{Matheus Hostert}
\email{mhostert@g.harvard.edu}
\affiliation{Department of Physics \& Laboratory for Particle Physics and Cosmology, Harvard University, Cambridge, MA 02138, USA}

\author{Maxim Pospelov}
\email{pospelov@umn.edu}
\affiliation{School of Physics and Astronomy, University of Minnesota, Minneapolis, MN 55455, USA}
\affiliation{William I. Fine Theoretical Physics Institute, School of Physics and Astronomy, University of Minnesota, Minneapolis, MN 55455, USA}

\date{\today}

\begin{abstract}
Current data on ultra-high-energy (UHE) cosmic rays suggest they are predominantly made of heavy nuclei. This indicates that the flux of neutrinos produced from proton collisions on the cosmic microwave background is small and hard to observe. Motivated by the recent extremely-high-energy muon event reported by KM3NeT, we explore the possibility of enhancing the energy-flux of cosmogenic neutrinos through nuclear photodisintegration in the presence of new physics. Specifically, we speculate that UHE neutrons may oscillate into a new state, dark (or mirror) neutron $n'$ that in turn decays injecting large amount of energy to neutrinos, $n\to n'\to \nu_\text{\tiny UHE}$. While this mechanism does not explain the tension between the KM3NeT event and null results from IceCube, it reconciles the experimental preference for a heavier cosmic ray composition with a large diffuse cosmogenic flux of UHE neutrinos.
\end{abstract}

\maketitle

\section{Introduction} 

The origin of ultra-high-energy (UHE) cosmic rays (CRs) is a long standing puzzle in astrophysics.
Below EeV energies, cosmic rays are significantly deflected by galactic and intergalactic magnetic fields, obscuring their origin.
High energy photons also lose information from their source due to energy losses to the cosmic microwave background (CMB) and the extragalactic background light (EBL).
Neutrinos, on the other hand, propagate undisturbed and can provide key input to identify individual UHECRs sources.
In addition to their pointing capabalities, the diffuse flux of UHE cosmogenic neutrinos provides some insight into the chemical composition of CRs~\cite{Ahlers:2009rf,2010JCAP...10..013K,Hooper:2004jc,Roulet:2012rv,vanVliet:2019nse}.

At energies of about $E_p > 50$~EeV, UHE protons experience a steep decrease in their mean-free-path ($\ell_{\rm mfp} \lesssim 50$~Mpc) thanks to resonant meson photo-production on the CMB, $p^+ \, \gamma_{\rm CMB} \to \Delta^+(1232) \to \pi^+ \, n \text{ or } \pi^0 \, p^+$.
This effect is referred to as the Greisen–Zatsepin–Kuzmin (GZK) cutoff~\cite{Greisen:1966jv,Zatsepin:1966jv} and imposes a propagation horizon for CR protons.
Unavoidably, such a cutoff also generates neutrinos~\cite{Berezinsky:1969erk,Hill:1983xs,Engel:2001hd,Hooper:2004jc,Takami:2007pp,Ahlers:2009rf}, which are free to propagate and can be detected at large neutrino telescopes like IceCube~\cite{IceCube:2016zyt} and KM3NeT~\cite{KM3Net:2016zxf}, and at cosmic ray experiments like Pierre Auger~\cite{PierreAuger:2019ens}.
Heavier nuclei are less effective at producing EeV neutrino fluxes since, for the same CR energy, the individual nucleon  energy is smaller. They are, however, a good source of UHE neutrons thanks to photo-disintegration processes on the CMB and EBL, such as $A \,\, \gamma \to A^*({\rm GDR}) \to (A -1) \,\, n$, where $A({\rm GDR})$ is a giant dipole resonance of the nucleus $A$.

GZK neutrinos from pion and muon decay carry about $\sim 5\%$ of parent proton energy while neutrinos from neutron $\beta$ decay are far less energetic, carrying typically $0.01\%$ of the parent proton energy thanks to the small $Q$-value of the reaction $n \to p^+ \, e^- \,\overline \nu_e$.
Therefore, neutrinos from nuclei photodisintegration have a much lower energy-flux and are much harder to detect when compared to their GZK analogues.
This fact combined with the apparent absence of a cosmogenic neutrino flux has been used by IceCube~\cite{IceCube:2016uab,IceCube:2025ezc} and Pierre Auger (PA)~\cite{PierreAuger:2019ens} to place upper limits on the proton fraction of CRs.

The PA and Telescope Array (TA) CR ground observatories also constrain the UHE CR composition through direct measurements of the extended air shower~\cite{PierreAuger:2014sui}.
Current data points to a heavy element CR composition at UHEs, with a fraction of protons that is large at EeV energies, but quickly drops to below $10\%$ above 10 EeV~\cite{PierreAuger:2022atd}.
The inference of this composition relies on an accurate modeling of the shower development, but despite large hadronic uncertainties, the available models are all consistent with a small proton fraction at UHEs~\cite{Ehlert:2023btz}.

Recently, KM3NeT reported the observation of an extremely high energy muon, $E_\mu \sim 120\text{ PeV}$, originated from just above the horizon, $\theta = 0.6^\circ \pm 1.5^\circ$~\cite{KM3NeT:2025npi}.
This event, dubbed KM3-230213A, is likely the highest energy neutrino ever observed and does not appear to be of atmospheric origin.
One explanation is that it is the result of a significant upward statistical fluctuation from the same diffuse, power-law flux observed by IceCube~\cite{IceCube:2020wum,IceCube:2020wum} when extrapolated to higher energies.
Under this hypothesis, assuming a single power law of $E_{\nu}^{-2}$, the KM3NeT event is consistent with a muon neutrino of energy $E_\nu = 220^{+570}_{-110}$~PeV and a flux of $E^2 \Phi (E) = 5.8^{+10.1}_{-3.7} \times 10^{-8} \, \text{GeV cm}^{-2}\text{s}^{-1}\text{sr}^{-1}$~\cite{KM3NeT:2025npi}.
The tension with the IceCube measurement, however, is large, at more than $(3 - 3.5)\sigma$~\cite{Li:2025tqf} (see also~\cite{KM3NeT:2025ccp}).
If the event comes from a transient source, this tension is reduced, but not below $2\sigma$.

Another hypothesis is that this event corresponds to a recovery or feature in the astrophysical neutrino flux from, e.g., the GZK process~\cite{KM3NeT:2025vut,Muzio:2025gbr}.
This cosmogenic hypothesis is also in $>3\sigma$ tension~\cite{Li:2025tqf} with IceCube~\cite{IceCube:2018fhm,IceCube:2025ezc} as well as with
PA searches~\cite{PierreAuger:2015ihf,PierreAuger:2019ens} thanks to their larger effective area in $E_\nu \gtrsim 100$~PeV energy region.
Nevertheless, given the neutrino energy interval estimated for the event and its consistency with the expected location of the cosmogenic neutrino energy-flux peak, in the rest of the paper, we proceed to entertain a cosmogenic origin for the event.

In this article, we show how new physics can lead to a large and detectable EeV cosmogenic neutrino flux even in the complete absence of protons in the CR spectrum.
We explore a``neutron portal" to new physics that leads to UHE neutrinos associated with CR neutrons instead of protons. 
The idea of a neutron portal is that free neutrons can decay or oscillate into some new beyond-Standard-the-Model states, that, in turn, decay with a significant amount of energy release.
Specifically, our scenario consists of an almost complete transformation of free neutrons into mirror/dark neutrons with a subsequent decay process, $n\to \nprime \to \nu \pi^\prime$ or $\nprime \to (p^\prime \to \pi^\prime \nu) \overline\nu_e^\prime e^\prime$, where $\pi^\prime$ is some new boson that can be identified with a mirror pion or another particle lighter than the neutron or proton.
Because of the larger $Q$ value of this mirror-baryon-number-violating reaction, the resulting neutrino can be much higher energy than its $n$ beta-decay analogue.
As we will see, these neutrinos could give rise to an observable cosmogenic flux that is compatible with the latest KM3NeT event, although still in tension with IceCube's null results.

This paper is divided as follows. 
In \cref{sec:neutronportal} we present the idea of a neutron portal to new physics in CRs and elaborate on an example based on a mirror sector. 
We describe our treatment of the time evolution of the neutron-mirror-neutron system in \cref{sec:evolution},  discuss potential UV origins of the models in \cref{sec:UV}, and discuss direct experimental limits in \cref{sec:constraints}. 
We then present and discuss our results for cosmogenic neutrino fluxes in \cref{sec:results} and conclude in \cref{sec:conclusions}. 
More details of our simulation of CR neutron production using \textsc{CRPropa3.2}~\cite{AlvesBatista:2022vem} are presented in \cref{app:cosmicrays}.

\begin{figure}[t]
    \centering
    \includegraphics[width=0.35\textwidth]{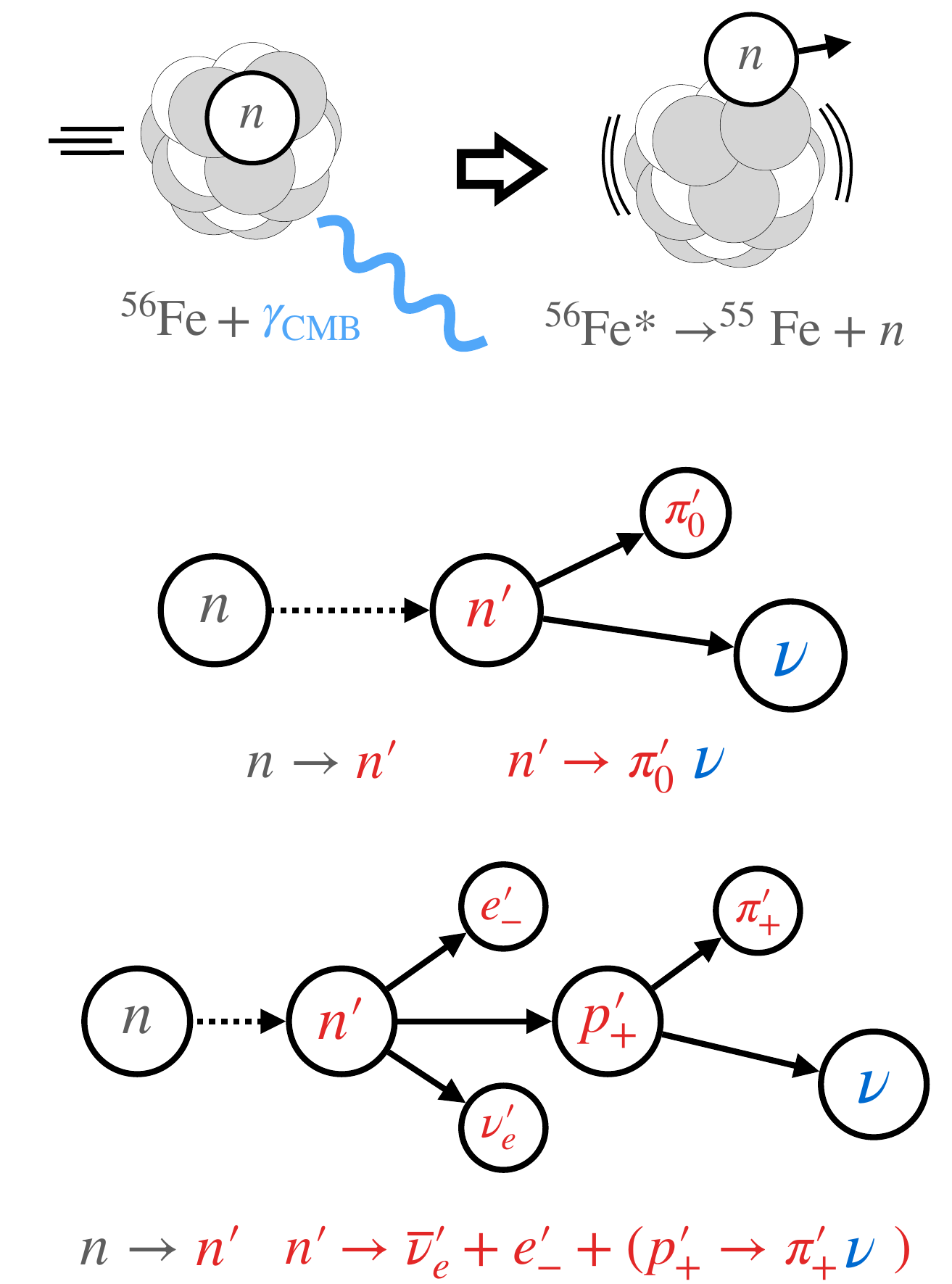}
    \caption{
    The photo-disintegration process of an iron UHE cosmic ray (top).
    We also show the $n \to n^\prime$ oscillation with two options for subsequent mirror-baryon decays: $\nprime \to \pi^\prime_0 \, \nu$ (middle) or $\nprime$ beta-decay followed by $p^\prime_+ \to \pi^\prime_+ \, \nu$ (bottom).
    \label{fig:diagram}
    }
\end{figure}

%%%%%%%%%%%%%%%%%%%%%%%%%%%%%%%%%%%%%%%%%%
\section{Neutron portal to UHE neutrinos}
\label{sec:neutronportal}

Neutrons are produced through photo-disintegration, when infrared-photons, including the CMB or EBL, excite the cosmic ray heavy nuclei, which subsequently decay into free nucleons and/or photons. They quickly escape their parent nucleus and propagate in the intergalactic medium (IGM) essentially as free particles.
This constitutes a unique probe of neutrons in conditions that cannot be reproduced in the lab: densities are extremely small and the magnetic fields, although ultimately unknown and uncertain, are indirectly constrained to be $10^{-15}~\text{G}\lesssim B_{\rm IGM} \lesssim 10^{-9}~\text{G}$~\cite{doi:10.1126/science.1184192,2011A&A...529A.144T,Pshirkov:2015tua}.

Upon propagation, free neutrons will decay with a lifetime of $E_n/m_n \times \tau_n \sim 10\text{ kpc} \times (E_n/\text{ EeV})$, producing neutrinos with an energy that is at most $E_{\bar\nu_e}/E_n \simeq Q_n/m_n \sim (0.782\text{ MeV})/(939\text{ MeV}) \sim 8 \times 10^{-4}$ smaller than that of the parent neutron thanks to the smallness of $Q_n = m_n - (m_p + m_e)$. Similarly, tritium, and heavier $\beta^\pm$-unstable nuclei can also be produced and decay to low-energy neutrinos.
While the protons produced in this process can also source neutrinos through the GZK process, a cosmic ray spectrum composed mainly of high-energy nuclei will be far less efficient at producing $\mathcal{O}(\text{EeV})$ neutrinos due to the smaller energy-per-nucleon and the kinematical threshold for $\pi^\pm$ production.

Therefore, a CR composition that is dominated by heavy nuclei at higher energies is unlikely to produce an observable cosmogenic neutrino flux. 
To overcome this limitation, we engineer an extension of the SM where the free neutrons produced from photo-disintegration oscillate into a dark baryon/mirror sector $n^\prime$, coincident in mass. Once such oscillations develop, $n^\prime$ can then decay back to UHE neutrinos that carry a larger fraction of the $n'$---and hence from $n$---in comparison to $\beta$-decays. We show that such transitions are still allowed by experimental constraints as the near vacuum conditions and extremely small magnetic fields of the intergalactic medium probes a unique environment where $n\to \nprime$ transitions can benefit from the near-degeneracy of the two neutral states.  

%%%%
\subsection{The mirror sector}

Our model for the mirror neutron scenario is given by the following low-energy Lagrangian:
\begin{equation}\label{eq:lagrangian}
    \mathscr{L} \supset \left(\epsilonmix \overline{n} \nprime + \text{ h.c.}\right) + m_n \overline n n + (m_n+\delta m) \overline\nprime \nprime,
\end{equation}
where $\epsilonmix$ is the mass mixing parameter, $m_n$ the neutron mass, and $\delta m$ is the mass splitting.  For $\delta m \ll \epsilon_{nn'}$, in vacuum, $n$ and $n'$ form two nearly degenerate states with masses $m_n \pm \epsilon$ and their mixing is maximized. In addition to (\ref{eq:lagrangian}), we assume further couplings of $\nprime$ that generate its decay, with total decay width $\Gamma_{\nprime}$. Given that $n$ and $\nprime$ mix we define the quantities $\Gamma_n$ and $\Gamma_\nprime$ in the limit of $\epsilonmix \to 0$.

An important observation for our scenario is that the cosmic neutrons propagate in a nearly perfect vacuum environment. Any residual medium effects alter their propagation, leading to a shift in the neutron self-energy $\Delta E_n$ and a suppression of the oscillatory behavior. In the intergalactic space, they receive far fewer collisions with other particles, even compared to the best ultra-cold neutron (UCN) bottle experiments in a lab. Below, we list the most likely sources of medium effects and estimate their contribution to $\Delta E_n$:
\begin{enumerate}
    \item The density of atoms in the intergalactic medium is completely negligible for generating a sizable contribution to $\Delta E_n$.

    \item In the presence of an intergalactic magnetic field, the neutron magnetic moment $\mu_n$ will induce an energy level splitting equal to $V_{\rm B_{IGM}} = \pm \gamma_n \vec{\mu} \cdot \vec{B}_{\rm IGM}$, depending on the neutron polarization. 
    Note that in the neutron's rest frame the value of the field $\gamma_n B_{\rm IGM}$ is significantly enhanced by the boost factor.
    Saturating the lower limit on the magnetic field, we find 
    \begin{equation} \label{eq:BIGM}
        |V_{\rm B_{IGM}}| \sim 6\times 10^{-18}\text{ eV}\times \left(\frac{\gamma_n}{10^{9}}\right)\left(\frac{B_{\rm IGM}}{10^{-15}\text{ G}}\right).
    \end{equation}
    \item EBL and CMB photons induce a medium-dependent magnetic and electric dipole moment for the neutron, shifting its energy according to the interaction potential $V_{\rm pol} = \pm \frac{\gamma_n^2}{2}\left(\alpha |\vec{E}_{\rm CMB}|^2 + \beta |\vec{B}_{\rm CMB}|^2\right)$, in the in the neutron rest frame. Taking the neutron polarizabilities to their measured values $\alpha + \beta \simeq 1.5 \times 10^{-3}$~fm$^{3}$, $T_{\rm CMB} \sim 2.3\times10^{-4}$~eV, and expressing electric and magnetic fields in terms of the  CMB energy density, $|\vec{E}_{\rm CMB}|^2 = |\vec{B}_{\rm CMB}|^2 = 4 \pi \rho_{\rm CMB} = 4\frac{\pi^3}{15} T_{\rm CMB}^4$ we find a negligible shift of
    \begin{equation}
        |V_{\rm pol}| \sim  2\times 10^{-23}\text{ eV}\times \left(\frac{\gamma_n}{10^{9}}\right)^2.
    \end{equation}
    \item Incoherent scattering on the CMB and EBL contributes to the imaginary part of $\Delta E_n$. 
    Assuming a mean free path of $\ell_{\rm mfp}\sim (1-100)$~Mpc for neutrons above GZK energies, the absorption rate in the neutron rest frame corresponds to 
    \begin{equation}
        \Gamma_{\rm GZK} \sim \frac{\gamma_n}{\ell_{\rm mfp}} \sim \left(6\times 10^{-21} - 6\times 10^{-23}\right)\text{ eV}.
    \end{equation}

    \item Mirror $\nprime$ state may also develop its own energy shift $\Delta E_{n'}$ due to, {\em e.g.}, interaction with dark radiation, dark matter and dark magnetic fields.  For large classes of dark matter models, the number density of dark matter particles is small. Furthermore, we know that the mirror radiation background should be smaller than that of the CMB~\cite{Berezhiani:2005hv,Berezhiani:2006je}. Thus, it is reasonable to expect this model-dependent contribution to be negligible as well.
    
\end{enumerate}
In such near-vacuum conditions, the initial neutron system can decay to SM products via the weak interactions, $n \to p^+ e^- \bar\nu_e$, or to mirror particles as they oscillate to $n'$. 
To induce a large cosmogenic flux, we will consider mirror-nucleon decays that violate mirror baryon number ($B'$) to produce UHE SM neutrinos. 
We focus on the following possibilities:
\begin{align}
    \label{eq:nprime_decay}
    &\quad n\to n^\prime \to \pi^\prime_0 \nu,
    \\
    \label{eq:pprime_decay}
    &\quad n\to n^\prime \to p^\prime_+ e^\prime_- \nu^\prime, \text{ with } p^\prime_+ \to \pi^\prime_+ \nu,
\end{align}
where primed particles are in the mirror sector and may also carry a mirror-electromagnetic charge denoted by their subscript.
Note that in both scenarios, a nucleon has decayed and transferred a significant fraction of its energy into light particles: a light boson ($\pi'_0$ or $\pi'_+$) and a SM neutrino $\nu$.
In \cref{eq:pprime_decay}, $E_{p^\prime}\sim E_n$ thanks to $m_{\nprime} \sim m_{p^\prime}$. In the ultra-relativistic regime, the final daughter neutrino energy is therefore $E_\nu \lesssim (1 - m_{\pi^\prime}^2/m_{N}^2) \times E_{N'}$ for the nucleon decay $N' \to \pi' \nu$.

%%%%%%%%%%%%%%%%%%%%%%%%%%%%%%%%
\subsection{Open quantum system approach of $n\to\nprime$ oscillations with decay}
\label{sec:evolution}

\begin{figure}[t]
    \centering
    \includegraphics[width=\linewidth]{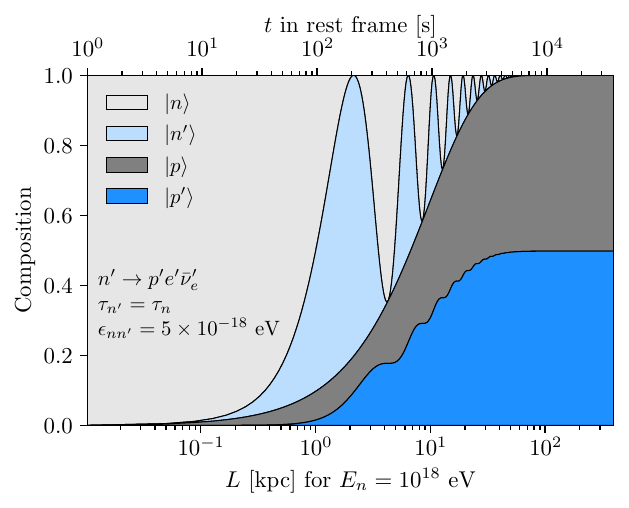}
    \caption{The time dependence of a system initialized as a neutron with $E = 10^{18}$~eV in terms of its fractional composition (stacked).
    Here, $\ket{p^\prime} = \ket{\pi'_0\nu}$ or $\ket{p^\prime} = \ket{p^\prime e^\prime \bar \nu_e^\prime}$ is the daughter state produced by $n^\prime$ decays.
    \label{fig:composition}
    }
\end{figure}

Particle decay in quantum mechanics is usually implemented following the Wigner-Weiskopff prescription~\cite{Weisskopf:1930ps}. Effectively, this amounts to changing the two-level Hamiltonian $H$ of the $n-\nprime$ system to a non-hermitian operator, $H \to H - i \Gamma/2$, where $\Gamma = \text{diag}(\Gamma_n, \Gamma_{\nprime})$, with $\Gamma_n$ ($\Gamma_{\nprime}$) the total decay width of $n$ ($\nprime$). This way, we get the standard exponential damping behavior when we time evolve the initial neutron system. Note, however, that this means that the probability of finding the state as a $n$ or $\nprime$ goes to zero at asymptotic times and is not conserved. This is because the daughter states are not included in the description.

A probability-preserving treatment is achieved by implementing decays from an open system perspective~\cite{Bertlmann:2006fn, Caban:2005ue}. For that, we extend our two-level $(n,\nprime)$ description to a four-dimensional system $(n,\nprime,p,\pprime)$ represented by the density matrix
\begin{equation}
    \rho(t) = \sum_{a, b} \rho_{a b}(t) \ket{a} \bra{b}, \quad a,b \in \{n,n,p,p^\prime\}
    \label{eq:state}
\end{equation}
with its eigenvalues interpreted as probabilities as its trace satisfies $\Tr(\rho(t)) = 1$ at any time. Without decay, the system's time evolution is driven by
\begin{equation}
    \pdv{\rho (t)}{t} =  -i[H,\rho(t)],
    \label{eq:time_evol_without_decay}
\end{equation}
where
\begin{equation}
    H = \sum_{a =n,n',p,p'}E_a \ketbra{a}{a} + \epsilonmix(\ketbra{n}{n'} + \ketbra{n'}{n}),
\end{equation}
is the Hamiltonian of the system.
For convenience of presentation, we label dark decay as $\nprime\to p'$, while in reality $\nprime$ can have a multitude of decay channels. 
Decay can be implemented, while keeping the trace normalization, by modifying Eq.~\eqref{eq:time_evol_without_decay} to 
\begin{align}
   \begin{split}
        \pdv{\rho (t)}{t} &=  -i[H,\rho(t)]\\
        &- \frac{1}{2}(\Pi^\dagger \Pi \rho(t) + \rho(t)\Pi^\dagger \Pi - 2 \Pi \rho(t) \Pi^\dagger),
   \end{split}
   \label{eq:time_evol_with_decay}
\end{align}
where
\begin{equation}
    \Pi = \sqrt{\Gamma_n} \ket{p} \bra{n} + \sqrt{\Gamma_{n'}} \ket{p^\prime} \bra{n^\prime},
    \label{eq:dissipation}
\end{equation}
is the operator associated with $n\to p$ and $n' \to p'$ transitions. Note that $\Tr(\Pi^\dagger \Pi \rho(t) + \rho(t)\Pi^\dagger \Pi - 2 \Pi \rho(t) \Pi^\dagger) = 0$, such that the trace is preserved throughout the time evolution. 

By working out the matrix multiplication in Eq.~\eqref{eq:time_evol_with_decay} and, as $\rho = \rho^\dagger$, the problem is reduced to solving 10 equations of motion with the initial condition $\rho(t=0) = \text{diag}(1,0,0,0)$, i.e., only neutrons. 
For instance, the dynamics of the neutron system is given  by
\begin{equation}
    \pdv{\rho_{nn}(t)}{t} = - \Gamma_n \rho_{nn}(t) + i \epsilon (\rho_{n \nprime}(t) -  \rho_{\nprime n}(t)).
\end{equation}
The time evolution of the off-diagonal entries, $\rho_{np}$ and $\rho_{pn}$, will depend only on themselves, and as they are zero at $t=0$ they vanish at all times. 

In \cref{fig:composition}, we show the time evolution of a free neutron, assuming $\delta m = \Delta E = 0$, $\Gamma_{n'} = \Gamma_n$, and taking $\epsilonmix = 10^{-18}$~eV. The probability of finding the initial neutron system as, for example, $p'$ can be computed as $\text{P}_{n \to \pprime}(t) = \Tr(\rho(t)\ketbra{p'}{p'})$. In the asymptotic limit $t \gg 1/\Gamma_{n(n')}$, with $\delta m = \Delta E = 0$, we find
\begin{equation}
        \text{P}_{n \to \pprime}(t\to \infty) = \frac{4 \Gamma_{n'} \epsilon^2}{(\Gamma_{n} + \Gamma_{n'})(\Gamma_n \Gamma_{n'} + 4 \epsilon^2 )},
        \label{eq:asymptotic_prob}
\end{equation}
therefore the transition $n \to p^\prime$---recalling that $\ket{p^\prime} = \ket{\nu_\alpha \pi^\prime}$ or $\ket{p^\prime} = \ket{p^\prime e^\prime \overline\nu^\prime_e}$---is largest when $\epsilon > \Gamma_{n'} >\Gamma$, as in this case Eq.~\eqref{eq:asymptotic_prob} reduces to 
\begin{equation}
    \text{P}_{n \to \pprime}(t\to \infty) = \frac{\Gamma_{n'}}{\Gamma_n + \Gamma_{n'}},
\end{equation}
which justifies the asymptotic behavior in Fig.~\ref{fig:composition}, as for $\Gamma_n = \Gamma_{n'}$ the probability is $1/2$. Therefore, in vacuum, a significant fraction of neutrons can convert into mirror sector particles. 

\begin{figure}[t!]
    \centering
    \includegraphics[width=\linewidth]{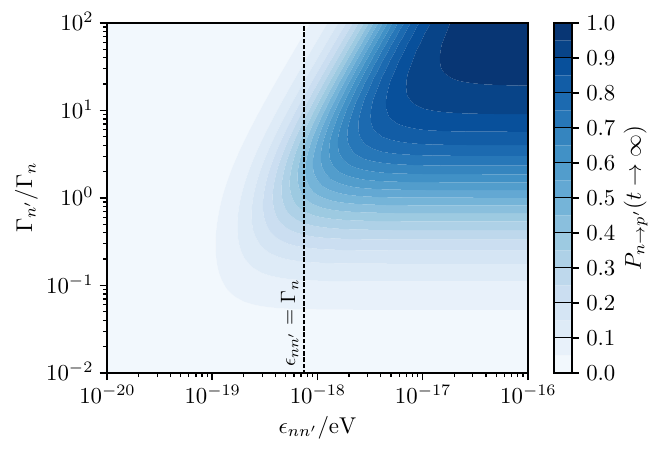}
    \caption{The time-asymptotic probability of $n \to n^\prime \to \nu \varphi$ transition as a function of the mass mixing parameter $\varepsilon_{n n^\prime}$ and the ratio of the $n^\prime$ to $n$ decay width.
    }
    \label{fig:asymptotic_prob}
\end{figure}

In Fig.~\ref{fig:asymptotic_prob} we show how to get a large $n \to n^\prime$ transition varying $\Gamma_{n'}$. The neutrons must oscillate before decaying, i.e., the mixing parameter must satisfy $\epsilon_{nn'}> \Gamma_n$. Hence, we require $\epsilonmix \sim \mathcal{O}(10^{-18})$~eV. This choice also evades lab constraints $\epsilonmix \ll 1/(352~\text{s}) = 1.87\times10^{-18}$~eV on the disappearance of ultra-cold neutrons (UCNs)~\cite{nEDM:2020ekj}. Once oscillations develop, the $n'$ decay width, $\Gamma_{n'}$, governs how much the mirror sector is populated. We note from Eq.~\eqref{eq:asymptotic_prob} that for $\Gamma_{n'} \Gamma_n > \epsilon_{nn'}^2$, the probability is suppressed. Therefore, in general the probability of $n\to\nprime\to p'$ is maximized if $\epsilon_{nn'}> \text{max}(\Gamma_n,\Gamma_{n'})$. 

Ultimately, our goal is to convert as many neutrons into UHE neutrinos as possible. This is done in two steps, first the neutrons oscillate $n \to n'$ and in sequence the mirror neutron decays into neutrinos, $n'\to \nu$. Next we introduce the dynamics of the mirror sector to show how the second step can be realized and discuss the relevant constraints.

%%%%%%%%%%%%%%%%%%%%%%%%%%%%%%%%%%%%%%%%%%%%%%%%
\subsection{UV completion}
\label{sec:UV}

The mixing parameter $\epsilonmix$  required for our argument is small and can be generated by new physics at energies far above the electroweak one.
For instance, as discussed in \cite{Berezhiani:2005hv,Berezhiani:2006je,Fornal:2020gto}, two scalar di-quarks $\Phi$ and $\Phi^\prime$, one colored under QCD and the other under QCD$^\prime$, and a singlet fermion $N$ that can be relatively light, can generate the mass mixing parameter at tree-level without being ruled out by direct searches.
In particular, for our mixing parameter and $\mathcal{O}(1)$ couplings between the quarks and $\Phi^{(\prime)}$,
\begin{equation}
\epsilon_{n n^\prime}  \sim 10^{-18} \, {\rm eV} \left( \frac{10 \,{\rm GeV}}{M_N} \right) \left( \frac{100\, {\rm TeV}}{M_{\Phi}}\right)^2
\left( \frac{100\, {\rm TeV}}{M_{\Phi^\prime}}\right)^2,
\end{equation}
with all particles outside of direct experimental reach.

As we show above, the lifetime of $\nprime$ should be close to that of the neutron, $0.1 \lesssim \Gamma_{n^\prime}/\Gamma_n \lesssim 10$, as otherwise the transition probability of $n \to n'\to \nu_\text{\tiny UHE}$ is too small to induce an observable cosmogenic neutrino flux.
Clearly, the process in \cref{eq:pprime_decay} would explain this coincidence, since for an exact mirror symmetry, $\Gamma(n \to p^+ e^- \overline\nu) = \Gamma(n' \to p'_+ e'_- \overline\nu')$.
In that case, for our cosmogenic neutrino flux to be observable, the mirror proton lifetime must be short enough for them to decay to neutrinos before reaching the Earth.
For a travel distance of $10$~Mpc and EeV mirror-protons energies, this requirement is $\tau_{p^\prime}/\tau_n \lesssim 10^{3}$, which is a loose criterion and provides more freedom from the model-building point of view.

The underlying physics behind the $B'$-violating nucleon decays in  \cref{eq:nprime_decay,eq:pprime_decay} can be very different from that of weak interactions and there may be no reason to expect that $\Gamma(n' \to \pi_0' \nu)$ or $\Gamma(p' \to \pi_+' \nu)$ be close to the weaker-than-Weak beta-decay rate of the neutron, $\Gamma_n \propto G_F^2 Q_n^5$.
In fact, given the larger energy release $Q_{N'} = m_{N'} - m_{\pi^\prime}$ in $B'$-violating decays, the effective operator responsible for mirror nucleon decays can arise from scales far above the electroweak one.

To illustrate our point, let us consider a specific model for the $B'$-violating decays.
We introduce a scalar leptoquark $S_1'$ (see, e.g., \cite{Dorsner:2016wpm} for a review) with charges $(\bar{\mathbf{3}},1, \frac{1}{3})$ under the mirrored gauge symmetry SU$(3)'\times$SU$(2)_L'\times$U$(1)'$.
The relevant couplings are
\begin{equation}
    \mathscr{L} \supset - \lambda_1 S_1'^* \left(Q'^T C Q'\right) - \lambda_2 S_1' \left(d'^T C \nu_R\right) + \mathrm{h.c.},
\end{equation}
where SU$(3)'$ and SU$(2)_L'$ indices are implicit, $Q'=\left(\begin{matrix}u'_L \\ d'_L \end{matrix}\right)$, $d' = d'_R$, and $C$ is the charge-conjugation matrix.
The only gauge-invariant operator involving SM leptons is the one above, which couples the leptoquark to $\nu_R$, a singlet under both the SM and the mirror sector gauge symmetries.
Note that mirror-nucleon decays produce antineutrinos in this picture.
If neutrinos are quasi-Dirac particles, the mirror-nucleon decays into $\overline\nu_R$, which in turn can oscillate to active SM antineutrinos via ultra-long-baseline oscillations.
This scenario would induce an additional suppression to the visible neutrino flux by the sterile-to-active oscillation probability.
Alternatively, if neutrinos are Majorana and $\nu_R$ is heavy and mixes with light neutrinos, then the mirror-nucleon decays can directly produce active neutrinos through mixing.
In this case, the decay rate is suppressed by a small mixing factor, but if $\nu_R$ is heavier than $n'$, then all mirror-nucleon decays will produce a visible neutrino flux.
The exact flavor composition of the neutrinos from mirror-baryon decay is in principle arbitrary, so for simplicity we assume flavor equipartition at Earth.

Integrating out the leptoquark, we find that the neutron decay rate is
\begin{equation}\label{eq:nprimedecay}
    \Gamma(n' \to \pi'_0 \, \overline{\nu}_R) = \frac{(m_n^2 - m_\pi^2)^2}{32 \pi} \frac{G_{n \nu}^2}{m_n} \left| \bra{\pi'_0} Q' Q' d'\ket{n'} \right|^2,
\end{equation}
where $G_{n \nu} \equiv \lambda_1 \lambda_2/M_{S_1'}^2$.
We can borrow the Lattice QCD calculations for proton decay to estimate $|\bra{\pi'_0} Q' Q' d'\ket{n'}| \simeq  0.15$~GeV$^2$~\cite{Yoo:2021gql}.
For the benchmark of $\Gamma(n^\prime \to \pi_0' \, \overline\nu) = \Gamma_n$, we get
\begin{equation}
    G_{n \nu} = 2 \times 10^{-12}~\text{ GeV}^{-2},
\end{equation}
which for $\lambda_1 \lambda_2 = \mathcal{O}(0.1)$, gives $M_{S_1'} \sim \mathcal{O}(200$~TeV).
The same model will also lead to $p' \to \pi'_+ \overline\nu$ decays with a decay rate analogous to that of \cref{eq:nprimedecay}.
Note that $\Gamma_{n'} = \Gamma(n'\to p'_+ e'_- \overline\nu') + \Gamma(n' \to \pi'_0  \overline\nu)$, so raising the leptoquark mass will recover the coincidence $\Gamma_n \simeq \Gamma_{n'}$ and facilitate the transition chain $n\to n' \to p' \to \overline\nu$ shown at the bottom of \cref{fig:diagram}.
Mirror-nucleon decays in this example of UV completion produce antineutrinos, however, neutrino and antineutrino cross sections are approximately the same at the energies we are interested in, so this difference has little consequence for the phenomenology.

Finally, we note that the tiny mass splittings $\delta m = m_n - m_{n'} \ll \epsilonmix\sim \mathcal{O}(10^{-18})$~eV considered here are fine-tuned from a theoretical point of view.
This is because the mirror symmetry cannot be exact in nature, as otherwise the number of degrees of freedom in the early Universe would be twice as large~\cite{Berezhiani:2005hv,Berezhiani:2006je}.
In our construction, this is also evident in the fact that the mirror baryon number $B'$ is violated while the SM baryon number $B$ remains conserved up to $n-n'$ mixing, which breaks $B - B'$.
As a consequence, the running of $\alpha_s$ and $\alpha_s^\prime$ differ and, therefore, so does the QCD scale, making the choice of $\delta m \ll \epsilonmix \sim \mathcal{O}(10^{-18}$~eV) in the IR a fine-tuned one~\cite{Fornal:2019eiw,Babu:2021mjg}.

\begin{figure*}[t]
    \centering
    \includegraphics[width=\textwidth]{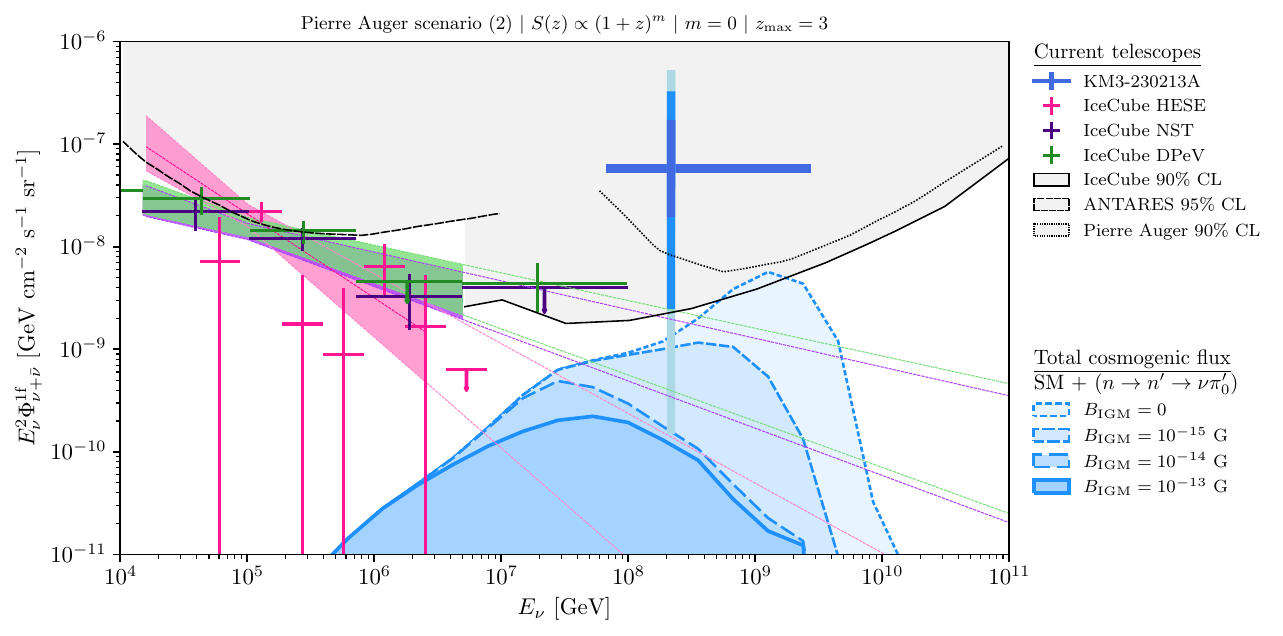}
    \caption{The single-flavor $\nu+\bar\nu$ flux from $n^\prime \to \pi'_0 \, \nu$ or $p^\prime_+ \to \pi'_+ \, \nu$ decay processes (blue lines) for $\epsilonmix = 10^{-18}$~eV, $\Gamma_n = \Gamma_{\nprime}$, and varying values of the intergalactic magnetic field $B_{\rm IGM}$.
    The data point in different shades of blue shows the $1\sigma$, $2\sigma$, and $3\sigma$ ranges of the neutrino flux to explain the KM3-230213A event at KM3NeT under the assumption of a clipped single-power law flux of constant $E^2 \Phi^{1{\rm f}}_{\nu + \overline{\nu}}$ in the range $72$~PeV $< E_\nu < 2.6$~EeV.
    Also shown are the existing limits on a diffuse cosmogenic flux from IceCube~\cite{IceCube:2018fhm,IceCube:2025ezc}, Pierre Auger~\cite{PierreAuger:2019ens}, and ANTARES~\cite{ANTARES:2024ihw}, as well as the measured astrophysical neutrino spectrum in three IceCube analyses: High-Energy Starting Events (HESE)~\cite{IceCube:2020wum} Northern Starting Tracks (NST)~\cite{Abbasi:2021qfz}, and the Downgoing PeV (DPeV) sample~\cite{Abbasi:2025bog}.
    }
    \label{fig:cosmo_flux_1}
\end{figure*}

\begin{figure*}[t]
    \centering
    \includegraphics[width=\textwidth]{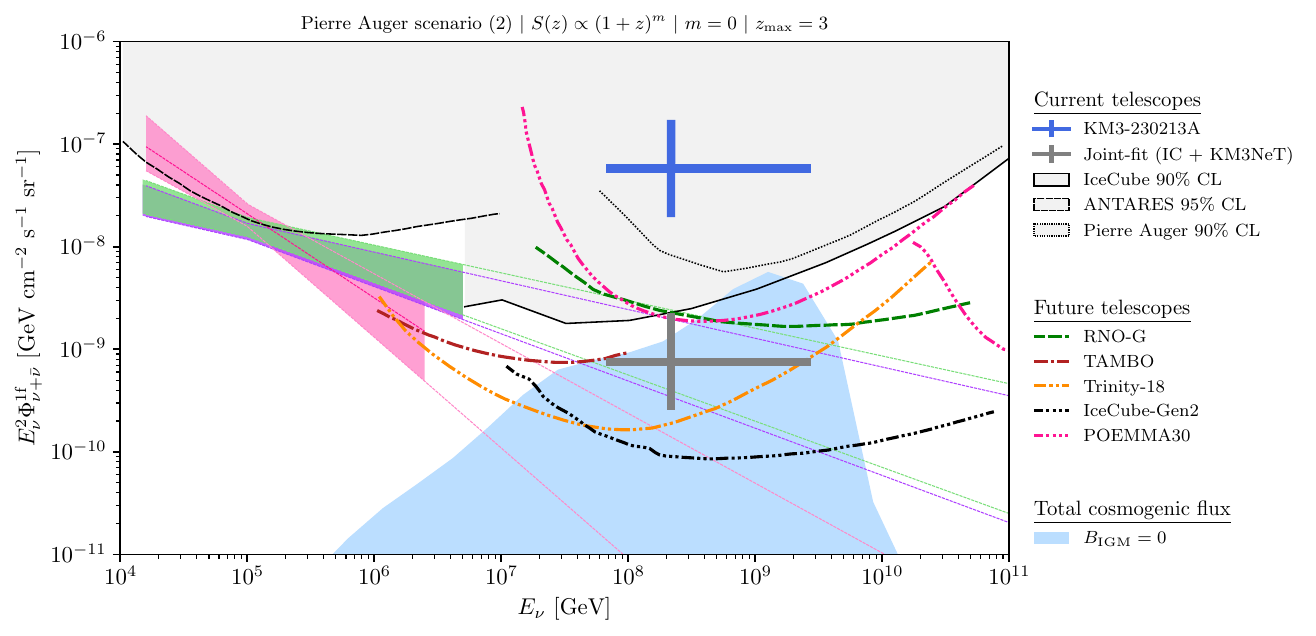}
    \caption{
    Same as \cref{fig:cosmo_flux_1}, but including the sensitivity of future radio experiments as different colors and dashed styles for RNO-G~\cite{RNO-G:2020rmc,RNO-G:2021hfx}, TAMBO~\cite{Romero-Wolf:2020pzh}, Trinity-18~\cite{2019arXiv190708732N}, POEMA30~\cite{POEMMA:2020ykm}, and IceCube-Gen2~\cite{IceCube-Gen2:2020qha,Meier:2024flg}.
    Also shown is the joint fit to IceCube and KM3NeT of \cite{KM3NeT:2025ccp} under the clipped single-power-law flux of constant $E^2\Phi^{1{\rm f}}_{\nu + \overline{\nu}}$.
    }
    \label{fig:cosmo_flux_1_future}
\end{figure*}
%%%%%%%%%%%%%%%%%%%%%%%%%%%%%
\subsection{Existing constraints}
\label{sec:constraints}

\emph{Disappearance of ultra cold neutrons:} The $n-n^\prime$ conversion causes additional disappearance of ultra-cold neutrons in the laboratory.
This can be constrained by:
i) measurements of the neutron lifetimes with UCNs such as in the UCN$\tau$ experiment~\cite{UCNt:2021pcg}, ii) direct searches for $n-n^\prime$ transitions with UCNs such as in the nEDM experiment at PSI~\cite{nEDM:2020ekj}.
While the former is performed in a large magnetic field setting, the latter exploits tiny magnetic fields, setting a limit on the mass mixing parameter of $\epsilonmix < 1/(352~\text{s}) = 1.87 \times 10^{-18}$~eV.
The setup corresponds to a neutron storage bottle subject to three magnetic field configurations: $B$, $-B$, and $B=0$, with runs at $B=10$~$\mu$T and $B=20$~$\mu$T.
Other dedicated searches for UCN disappearance have also reported null results, but are not sensitive to the extreme degeneracy we consider here~\cite{PSI:Ban:2007tp,Serebrov:2008her,PSIlike:Serebrov:2009zz,PSI:Altarev:2009tg,Berezhiani:globalexpt:2017jkn,Gonzalez:2024dba}.
UCN disappearance is sensitive to $1-P_{n \to n}$ and so are independent of $\Gamma^\prime/\Gamma$ and the details of the invisible mirror-neutron decay.

\emph{Nuclear stability:} Nuclear stability in the SM is ensured by the small mass difference between the neutron and the proton.
Oscillations of $n\to \nprime$ with subsequent large-energy-release $\nprime$ decay can spoil this delicate balance.
Searches for invisible neutron decay in $^{12}$C and $^{16}$O at underground detectors such as KamLAND~\cite{KamLAND:2005pen}, Borexino~\cite{Borexino:2003igu}, and SNO~\cite{SNO:2003lol,SNO:2018ydj} are the most sensitive to these effects.
Specifically, SNO+ sets the stringiest limit of $\tau(n \to \text{inv}) > 9 \times 10^{29}$~years~\cite{SNO:2022trz}.

For the significant degeneracy between $n$ and $\nprime$ masses considered here, the $n\to \nprime$ transition in nuclei is extremely rare thanks to the large nuclear potential felt by the neutron.
While in vacuum the mixing angle of the two-state system is maximal, $\theta_{n\nprime} \sim \pi/4$ for $\delta m \to 0$, in nuclei, it is suppressed by a strong hierarchy of scales,
\begin{equation}\label{eq:nnprimemixing}
    \theta^{\text{nuclei}}_{n\nprime} \sim 10^{-25} \times \left(\frac{\epsilonmix}{10^{-18} \text{ eV}}\right) \left(\frac{10\text{ MeV}}{\Delta E_n}\right).
\end{equation}
For instance, taking $\tau_{\nprime} = \tau_n$, we can estimate
\begin{equation}
    \tau({}^{16}{\rm O} \to {}^{15}{\rm O}^{*} + \text{inv}) \sim \frac{\tau_n}{8 (\theta_{n\nprime}^{\rm Oxygen})^2} \sim \mathcal{O}(10^{44}~\text{years}),
\end{equation}
far beyond experimental limits.

Protons can also decay via $p^+ \to \pi^+ (n\to \nprime \to \text{inv})$ if the final states of the $\nprime$ decay are lighter than $m_p - m_\pi$.
Searches for the analogous two-body decay channel to a pion and a neutrino constrain $\tau(p^+ \to \pi^+ \nu)<3.9 \times 10^{32}$~years~\cite{Super-Kamiokande:2013rwg}.
Because proton decay proceeds through an off-shell neutron oscillation $n\to n^\prime$, the virtuality of the neutron suppresses the oscillation, where now $\Delta E$ in \cref{eq:nnprimemixing} is approximately $\Delta E \sim 100$~MeV.
Therefore, both nuclear and proton stability limits do not rule out the parameter space considered here.

\emph{Neutron stars:} One of the strongest limits on the $\epsilonmix$ mixing arises from considerations of the late time neutron star heating, as continuing transfer of $n$ to $\nprime$ creates vacancies on the Fermi sea of nucleons, that are filled with $O(E_F)$ energy release \cite{Goldman:2019dbq,Berezhiani:2020zck,McKeen:2020oyr,McKeen:2021jbh,Goldman:2022rth}. We note that in our scenario, the corresponding limits can be somewhat stronger. This is because subsequent decay of $\nprime$ to neutrinos will deposit up to $m_n/2$ energy due to the neutrino energy loss inside the star. Therefore the constraints on $\epsilonmix$ can be strengthened by a factor of a $O((m_\nprime/(2E_F)))^{1/2}\sim\,{\rm few}$, compared to those derived in {\em e.g.} Refs.~\cite{McKeen:2021jbh,Hostert:2022ntu}. This strengthening of the bounds can be avoided if the birth of the SM neutrino is preceded by a light dark state, such as $\nu_R$, that escape the star before decaying or oscillating into ordinary neutrinos.

\begin{figure*}[t]
    \centering
    \includegraphics[width=0.49\textwidth]{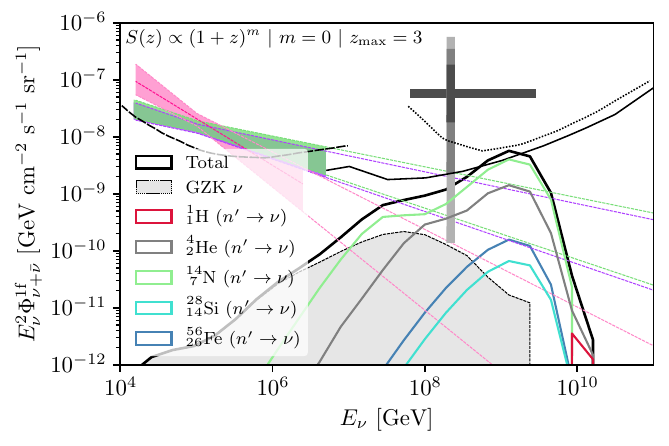}
    \includegraphics[width=0.49\textwidth]{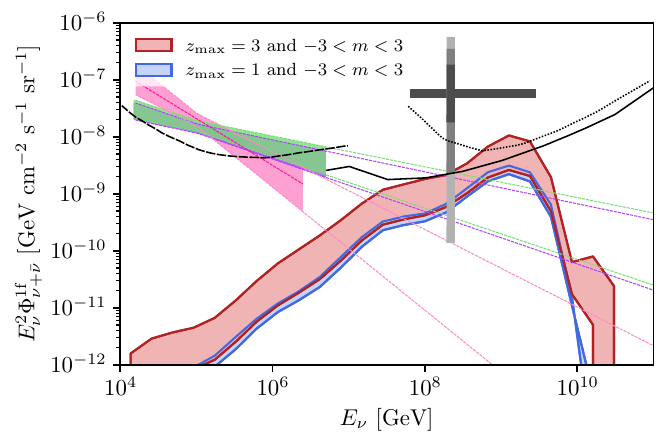}
    \caption{    
    Left) the cosmogenic neutrino flux of \cref{fig:cosmo_flux_1} with $B_{\rm IGM} = 0$ separated into the primary CR components of our benchmark scenario (see \cref{app:cosmicrays}).
    We show each individual flux from new physics (colored solid lines, not stacked), the total cosmogenic flux from standard GZK processes (dashed grey), as well as the total cosmogenic flux (black solid line).
    Right) the total cosmogenic neutrino flux for $B_{\rm IGM} = 0$ for various assumptions on the source evolution. The flux envelope is defined by the two extreme values of $-3 < m < 3$ and is shown for fixed $z_{\rm max} = 1$ (blue) and $z_{\rm max} = 3$ (red).
    }
    \label{fig:cosmo_flux_2}
\end{figure*}

%%%%%%%%%%%%%%%%%%%%%%%%%%%%%
\section{Results and discussion}
\label{sec:results}

In \cref{fig:cosmo_flux_1}, we show the cosmogenic neutrino plus antineutrino flux per flavor in our cosmic ray benchmark model.
We assume a flavor equipartition at Earth, $(\nu_e:\nu_\mu:\nu_\tau)=(1:1:1)$.
The various blue curves include neutrinos from ordinary pion and neutron decays, as well as from $B^\prime$-violating decays for various assumptions of the IGM magnetic field $B_{\rm IGM}$.
The latter modifies the $n\to n'$ oscillation probability by effectively suppressing the $n-n'$ mixing angle due to the energy splitting in \cref{eq:BIGM}.
For $B_{\rm IGM} = 10^{-13}$\,G, the oscillation is suppressed enough that the neutrino flux from SM processes, such as $\pi^\pm$ decay, dominates.

We also show the segmented-flux data points from the astrophysical neutrino analyses at IceCube, including the following samples: High-Energy Starting Events (HESE)~\cite{IceCube:2020wum}, the Northern Starting Tracks (NST)~\cite{Abbasi:2021qfz}, and the Downgoing PeV (DPeV)~\cite{Abbasi:2025bog}, each under the assumption of a single, unbroken power law.
The best-fit and the uncertainty band of the power law flux are shown as a solid line and shaded region within the relevant energy region of each analysis.
The extrapolation to high energies is shown as a continuation of the latter in dashed lines and lighter tones.

The diffuse flux needed to explain the one event from KM3NeT~\cite{KM3NeT:2025npi} under the assumption of an $E^{-2}$ power-law flux is shown as a blue datapoint, including the $1\sigma,2\sigma,3\sigma$ vertical error bars as different shades of blue. 
The horizontal error bar stands for the central $90\%$~CL interval on the reconstructed neutrino energy of the event.
The diffuse fluxes excluded by IceCube~\cite{IceCube:2018fhm,IceCube:2025ezc}, Pierre-Auger~\cite{PierreAuger:2019ens}, and ANTARES~\cite{ANTARES:2024ihw} are shown as grey shaded regions.
In particular, the search in \cite{IceCube:2025ezc} is in strong tension with a cosmogenic neutrino flux explanation of the KM3NeT event, as explained in \cite{Li:2025tqf}.

\Cref{fig:cosmo_flux_1_future} shows the future sensitivity of radio telescopes such as RNO-G~\cite{RNO-G:2020rmc,RNO-G:2021hfx}, TAMBO~\cite{Romero-Wolf:2020pzh}, Trinity-18~\cite{2019arXiv190708732N}, and POEMMA30~\cite{POEMMA:2020ykm}.
Eventually, IceCube-Gen2~\cite{IceCube-Gen2:2020qha,Meier:2024flg} will improve IceCube's sensitivity to even lower cosmogenic fluxes.
These could help determine the presence and origin of a cosmogenic neutrino flux.
With more PA and TA data, one can even speculate on an even stronger tension between the CR composition and cosmogenic models, motivating alternative cosmogenic neutrino scenarios, such as strong in-source radiation fields, or new physics constructions such as our own.

The left panel of \cref{fig:cosmo_flux_2} shows the total cosmogenic flux separated into the primary CR composition.
The GZK process on protons is responsible for a few UHE neutrons that can oscillate to mirror neutrons at the highest energies, visible as a red peak.
The vast majority of new-physics-induced neutrinos come from nuclear photo-disintegration at lower energies.
Due to the dominance of nitrogen in the CR population for our benchmark model (see \cref{app:cosmicrays}), this component dominates the cosmogenic flux, suggesting that most CR neutrons arise from intermediate mass nuclei.
Neutrinos from standard photo-meson production and neutron beta decay are shown in grey.
Due to the limited sample size of our simulation, this flux is only visible up to $E_\nu \sim 2\times 10^9$~GeV.

The right panel of \cref{fig:cosmo_flux_2} shows the total cosmogenic neutrino flux in the presence of new physics for varying assumptions on the CR source evolution parameters, namely $m$ and $z_{\rm max}$.
We assume $B_{\rm IGM}=0$ in this case as well.
As expected, for strongly positive source evolutions, the neutrino flux can be enhanced, especially for larger CR horizons, $z_{\rm max}$.
As we vary the source evolution, we neglect the fact that for different $m$ and $z_{\rm max}$, the primary CR composition fractions could change in order to accommodate the PA data on Earth.

For the most optimistic source evolution of $m=3$, the expected number of events at KM3NeT for 335 days of exposure is between $0.14$ and $0.016$ for $B_{\rm IGM}$ in the range of $0$ to $10^{-13}$~G.
The probability of these fluxes producing a single event at KM3NeT is then $12\%$ and $1.4\%$.
The corresponding range of events in the IceCube analysis of Ref.~\cite{IceCube:2025ezc} is between $10 - 0.90$ events, further illustrating the tension between the two experiments under a diffuse flux interpretation of KM3-230213A.
These rates should also be contrasted with the expected cosmogenic neutrino rate of our CR benchmark model without new physics, which at KM3NeT and IceCube is $0.015$ and $0.8$, respectively.
Therefore, in this CR model, the novel neutron decays can enhance the expected cosmogenic rate by over an order of magnitude.
The largest uncertainties in this prediction is associated with the CR model and the exact composition of cosmic rays.
Indeed, the number of neutrons in the CR spectrum is ultimately unknown and is hard to directly constrain due to its short-lifetime.
The heavier the composition of UHE CRs at the sources, the more neutrons are produced and the larger the potential new physics flux can be~\footnote{In fact, some authors speculate on the presence of ultra-heavy nuclei~\cite{Zhang:2024sjp}, which, while rare, can be effective sources of photodisintegration neutrons.}.

We emphasize that a low CR proton fraction on Earth does not necessarily imply an unobservable cosmogenic neutrino flux since strong radiation fields around UHECR sources can be effective proton absorbers and neutrino emitters~\cite{Unger:2015laa,Muzio:2021zud} (for a recent review, see~\cite{Muzio:2025xen}). 
This is the case for high-luminosity Active Galactic Nuclei (AGN)~\cite{PhysRevLett.126.191101,10.1093/mnras/stad2740}, where the photon density is high enough for trapped UHECRs to undergo photodisintegration and photopion production, boosting the neutrino emission and suppressing the observable UHE CR proton fraction.
In addition, it has been pointed out that PA data does not exclude a proton fraction of $\mathcal{O}(10\%)$ in certain CR evolution models~\cite{Muzio:2025gbr}.
Therefore, disentangling a new physics explanation from more mundane astrophysical effects is challenging without complementary direct evidence for the new physics in the laboratory. 
In this case, improving limits on the disappearance of ultra-cold neutrons to timescales above the neutron lifetime would be the most effective way to constrain our scenario.

The CR composition can also be constrained by a cosmogenic UHE \emph{photon} flux~\cite{Hooper:2010ze}, which, similarly to neutrinos, is also enhanced for a pure proton composition.
In this context, a gamma ray follow up to the KM3NeT event is already constrained but may still be on the way to reveal the original source of the CRs~\cite{Fang:2025nzg}.
In our model, it is possible that if CR composition of the source has sufficiently small proton fraction, then a gamma-ray coincidence is not necessarily present.
If observed, however, one can entertain the possibility of $n' \to \pi^0 \nu \to \gamma \gamma \nu$ decays, which would provide a strong correlation between gamma rays and neutrinos.

Furthermore, the neutron portal can be explored in more ways in order to enhance or mimic a cosmogenic neutrino flux.
For instance, our neutron portal may produce dark particles that interact with ordinary matter and fake the signal of cosmogenic neutrinos. A dark particle $\chi$ produced in $n^{(\prime)} \to \chi \, \gamma$ decays can upscatter into a heavier state $\chi^*$ as it crosses the Earth atmosphere, subsequently decaying to charged leptons. This process is analogous to neutrino scattering, but is now associated with two length scales: the upscattering length of $\chi$ on SM nucleons and the decay length of $\chi^*$ back to SM particles. 
This scenario is akin to the one discussed in \cite{Bertolez-Martinez:2023scp}, from which we can conclude that for $\sigma_{\chi N \to \chi^* N} \sim\mathcal{O}(10^{-32}$~cm$^2)$ and $\tau_{\chi^*} \gtrsim \mathcal{O}(10^{-4}\text{ s})$ at EeV energies, one could also explain the anomalous events in ANITA-V~\cite{ANITA:2020gmv,ANITA:2021xxh}.
In these models, the flux of new particles $\chi$ should be similar to that of UHE CRs.
Our neutron portal provides a natural explanation of this coincidence.

%%%%%%%%%%%%%%%%%%%%%%%%%%%%%%%%%%%%%%%%%%%%%%%%%%%%%%%%%%%%%%%%
\section{Conclusions}
\label{sec:conclusions}

We presented a new physics portal to UHE cosmogenic neutrinos based on oscillations between free neutrons and mirror-neutrons in the intergalactic medium. 
While cosmogenic neutrinos are usually associated with a large CR proton fraction, we showed that a UHE CR spectrum made exclusively of heavy nuclei can lead to a large and observable cosmogenic neutrino flux on Earth in the presence of new physics. 
This mechanism is independent of other neutrino production mechanisms in the sources and relies solely on the unavoidable disintegration of UHE nuclei on the cosmic microwave and infrared background permeating the intergalactic medium.
In these unique conditions, neutrons behave almost like free particles and can undergo oscillations into a degenerate state in the mirror or dark sector provided intergalactic magnetic fields are sufficiently small.
As oscillations develop, mirror-baryon-number-violating decays of mirror neutrons or protons can release a significant fraction of their energy to daughter neutrinos---unlike ordinary beta decay, where neutrinos carry only an $\mathcal{O}(10^{-4})$ fraction of the parent energy.

For oscillations to efficiently populate the mirror sector, the $n-n'$ mixing must be of order $\epsilon_{nn'} \gtrsim \Gamma_n \simeq 7.5\times 10^{-19}$~eV, otherwise, oscillations would be cut short by neutron decay. 
This can be directly tested via UCN disappearance searches, offering a way to probe this model that is independent of UHE CR composition measurements, which are subject to significant hadronic and source uncertainties. 
Using a simplified CR source model, we showed that our scenario accommodates the preference for a heavier CR nuclei composition while still yielding an observable flux of UHE neutrinos in the case of a $B_{\rm IGM} \sim \mathcal{O}(10^{-15}~{\rm G})$. 
Like other potential new physics scenarios for an enhanced cosmogenic neutrino flux~\cite{Muzio:2025gbr,Boccia:2025hpm}, it may explain the KM3-230213A event but it would not explain the lack of observation of UHE cosmogenic neutrinos at IceCube (for recent new physics proposal to explain this tension, see~\cite{Brdar:2025azm}).

More data from IceCube and KM3NeT, as well as from future neutrino telescopes, can help determine if a diffuse cosmogenic neutrino flux has indeed been observed.
At the same time, more data from CR observatories, such as with the latest PA upgrade AugerPrime~\cite{PierreAuger:2016qzd}, can teach us more about the composition of UHE CRs.
Before then, we encourage experimental searches for the disappearance of UCN in near-vacuum conditions that aim to reproduce the conditions of intergalactic space.
This can be achieved, for example, with larger UCN storage vessels or better precision on the measurement of their disappearance, targeting oscillation times somewhat longer than the neutron lifetime.
Other interesting future directions to consider include investigating the production of other visible particles in mirror-baryon decays, such as in $n' \to \nu \pi^0$.
In that case, we expect a strong correlation between UHE neutrino and gamma-ray fluxes, as well as visible mirror-baryon-number-violating decays in the lab.

%%%%%%%%%%%%%%%%%%%%%%%%%%%%%%%%%%%%%%%%%%%%%%%%%
\section*{Acknowledgements} 
We thank Pedro Machado and Ting Cheng for discussions on KM3NeT.
We acknowledge support from the Simons Foundation Targeted Grant 920184 to the Fine Theoretical Physics Institute. 
MP is supported in part by the DOE grant DE-SC0011842.
The work of MH is supported by the Neutrino Theory Network Program Grant \#DE-AC02-07CHI11359 and the US DOE Award \#DE-SC0020250. G.F.S.A. received full financial support from the São Paulo Research Foundation (FAPESP) through the following contracts No. 2022/10894-8 and No. 2020/08096-0.

\appendix

%%%%%%%%%%%%%%%%%%%%%%%%%%%%%%%%%%%%%%%%%%%
\section{Simulating UHE CRs}
\label{app:cosmicrays}

\begin{figure}[t]
    \centering
    \includegraphics[width=0.49\textwidth]{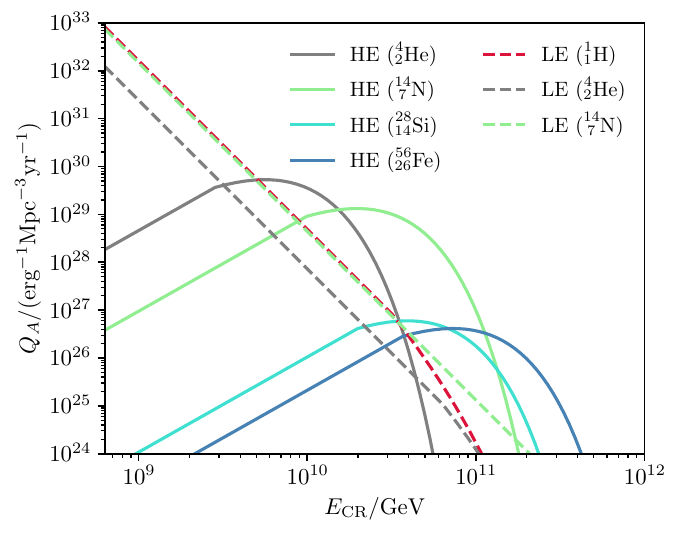}
    \caption{The injected spectrum of CRs at the sources following our CR model benchmark (scenario (2) of PA~\cite{PierreAuger:2022atd}).
    The low-energy (LE) and high-energy (HE) components are shown separately for various CR chemical compositions at the sources.
    \label{fig:source_spectra}}
\end{figure}

To simulate the cosmic ray evolution through the universe and estimate the resulting cosmogenic neutrino flux from the neutron portal to new physics, we make use of \textsc{CRPropa3.2}~\cite{AlvesBatista:2022vem}.
Our modeling of UHE CRs follows the latest Pierre Auger analysis~\cite{PierreAuger:2022atd} closely, where CR sources are assumed to be continuously distributed in redshift $z$ following some parametrization $S(z) = (1+z)^{m}$, for an integer $m$.
\textsc{CRPropa} uses the TALYS~\cite{Koning:2005ezu} for nuclear photodisintegration reactions and we use the EBL model from~\cite{Gilmore:2011ks}.
Cosmic rays are propagated in a one-dimensional simulation neglecting magnetic fields, which is a good approximation for the EeV energies considered here.

The energy injection spectrum for our benchmark CR scenario is shown in \cref{fig:source_spectra}, separating the CRs into low-energy (LE) and high-energy (HE) populations.
This corresponds to scenario (2) of Table 1 in Ref.~\cite{PierreAuger:2022atd}.
We find similar qualitative results for scenario (1).
The element fractions, luminosity, and maximum rigidity were fit to data in \cite{PierreAuger:2022atd}, where a good fit to composition and energy spectrum was observed.
The injection spectrum for a CR population of a nucleus of mass number $A$ and atomic number $Z$ was parameterized according to
\begin{equation}
    Q_{A}(E) = Q_{A}^0 \left(\frac{E}{E_0}\right)^{\gamma}\times 
    \begin{cases}
    1 \text{ if } E < Z R_{\rm max},
    \\
    e^{\left(1 - \frac{E}{ZR_A}\right)} \text{ if } E \geq Z R_{\rm max},
    \end{cases}
\end{equation}
with the total spectrum given by $Q = \sum_A Q_{A}$.
Here, $\gamma$ is the spectral index and $R_{\rm max}$ the maximum rigidity for a given CR population.
The total energy injected is then given by $J_A = \int_{E_0}^{\infty} E Q_A(E) \dd E$, with element fractions $I_A = J_A/\left(\sum_A J_A\right)$.
We neglect any galactic CR component.
The observed and propagated CR spectrum is then normalized to PA data at $E_{\rm CR} = 10^{17.8}$~eV, which then determines the neutrino fluxes on Earth.
We assume that neutron disappearance does not impact the fit to cosmic ray data.
This approximation is justified if protons from the decay of neutrons produced by photodisintegration or by the GZK process do not contribute significantly to the CR spectrum, which we find to be the case.

To obtain the new flux of neutrinos from $B'$-violating decays, we select events with neutrinos emitted in neutron beta decay and modify their properties accordingly.
The differential decay rate is assumed to be proportional to $(1 - E_\nu/E_n)$.
We show our results for the case $n' \to \nu \pi'_0$, but those for $p'_+ \to \nu \pi'_+$ will be similar since $p'$ takes most of the $n'$ energy in the mirror beta decay process.

\bibliographystyle{apsrev4-1}
\bibliography{main}
  
\end{document}